\documentstyle[aps,preprint,floats,tighten]{revtex}
\begin{document}

\preprint{\vbox{\hbox{UM-TH-95-27}
                \hbox{SU-4240-623}
                \hbox{EFI-95-71}
		\hbox{hep-ph/9511317}
		}
}

\title{Low-Energy Supersymmetry Breaking and Fermion Mass Hierarchies}

\author{Tony Gherghetta\footnote{tgher@umich.edu}}
\address{Department of Physics, University of Michigan,
	Ann Arbor, Michigan~~48109-1120}
\author{Gerard Jungman\footnote{jungman@npac.syr.edu}}
\address{Department of Physics, Syracuse University,
	Syracuse, New York~~13244}
\author{Erich Poppitz\footnote{epoppitz@yukawa.uchicago.edu}}
\address{Enrico Fermi Institute,
	5640 S. Ellis Ave.,
	Chicago, Illinois~~60637}

\date{November, 1995}

\maketitle

\begin{abstract}
In models with low-energy supersymmetry breaking, an anomalous
Abelian horizontal gauge symmetry can simultaneously explain
the fermion mass hierarchy and the values of the $\mu$ and $B$
terms. We construct an explicit model where the anomaly is cancelled
by the Green-Schwarz mechanism at the string scale. We show that with
our charge assignments, the breaking of the horizontal symmetry
generates the correct order of magnitude and correct hierarchy for
all Yukawa couplings.
\end{abstract}


\newpage

\section{Introduction}

Models of dynamical low-energy supersymmetry (SUSY) breaking
\cite{dnns,dn,dns} can provide a predictive and testable framework
for weak-scale supersymmetry.
In these models all mass scales arise via dimensional transmutation;
the soft SUSY-breaking parameters are calculable in terms of only a few
parameters. These low-energy SUSY-breaking models solve in a natural way
the flavour problem
of  hidden sector supergravity theories---the soft masses of the
quarks and leptons are proportional to their gauge quantum numbers and are
therefore flavour neutral.  At the weak scale these
models resemble the minimal supersymmetric
standard model, with a constrained parameter space.

A class of such models has been considered recently \cite{dnns}.
One difficulty with these models is the generation of appropriate
$\mu$ and $B$ terms at low energies, without significant fine-tuning.
In Ref.~\cite{dnns} this problem is circumvented by the introduction
of an extra singlet field with higher order non-renormalizable
couplings to the SUSY-breaking sector, which can provide a mechanism
for generating a $\mu$ term of appropriate magnitude.

In this paper we will illustrate another method for obtaining
effective couplings of the right magnitude to generate
acceptable $\mu$ and $B$ terms, without explicit fine-tuning.
As an added bonus, the same mechanism can generate the fermion mass
hierarchy, as has been recently explored elsewhere
\cite{ibanez,binetruy,dudas,jain,nir,rz}.
We propose to introduce an anomalous, generation dependent,
$U(1)_X$ gauge symmetry, under which the quark and lepton superfields
of the supersymmetric standard model transform nontrivially. In addition
the ``messenger'' singlet field responsible for communicating the
SUSY breaking is charged under the $U(1)_X$ symmetry.
The anomaly is cancelled by the Green-Schwarz mechanism. When the
$U(1)_X$ symmetry is broken, the necessary Yukawa couplings are induced.
The charge assignments of the various fields control
the induced coupling hierarchy in a manner similar to that of
the  Froggatt-Nielsen approach \cite{fn}.

The outline of this paper is as follows. We begin in Sec. 2 with a
brief review of the sector responsible for communicating SUSY breaking
to the SUSY standard model. We explain the difficulty in obtaining
$B$ and $\mu$ terms of order the electroweak scale without recourse to
significant
fine tuning or to the introduction of additional structure in the Higgs
sector. In Sec. 3 we outline the proposed solution of the fine-tuning
problem via an anomalous horizontal symmetry and state the conditions
necessary for the Green-Schwarz cancellation of the anomaly. In Sec. 4
we give an existence proof for models of this type, by presenting
explicit solutions of the Green-Schwarz anomaly cancellation
conditions. The given charge assignments explain
the fermion mass hierarchy and the generation of
acceptable $\mu$ and $B$ terms. Sec. 5 contains a summary of our results.

\section{Low-Energy SUSY-Breaking Model}

We will consider the model of low-energy SUSY breaking introduced
in Ref.~\cite{dnns}. For our purposes the full structure of the
SUSY-breaking model will be inessential. We will only
consider the sector responsible for communicating SUSY breaking to the
SUSY standard model (``messenger sector''). This sector consists of a pair
of charged (under ``messenger hypercharge'') superfields $\phi_\pm$,
a gauge singlet superfield $X$, and a pair of additional vectorlike
quark and lepton superfields, the ``messenger'' quarks $q, \bar{q}$,
and leptons $l, \bar{l}$.
The superpotential of the messenger sector is given by
\begin{equation}
	\label{wsb}
	W_M = k_1 X \phi_+\phi_- + k_2 X \bar{l}l + k_3 X \bar{q}q
	+ {1\over 3} \lambda X^3.
	\label{WSB}
\end{equation}
Due to their messenger hypercharge
interaction with the ``supercolor'' sector (the sector that
dynamically breaks SUSY), the $\phi_{\pm}$ fields acquire a negative mass
squared, $(-m_\phi^2)$, related to the scale of dynamical SUSY
breaking \cite{dns}.
The scalar potential of the messenger sector is
\begin{equation}
	\label{VSB}
	V_M = - m_\phi^2 (|\phi_+|^2 + |\phi_-|^2) +
	\sum_{\varphi} |{\partial W \over \partial \varphi}|^2~,
\end{equation}
where $\varphi$ denotes all the fields in the messenger sector.
A minimum of Eq.~(\ref{VSB}) with nonvanishing vacuum expectation values
({\it vev}s) for
$\phi_{\pm}, X$, and the $F$-component $F_X$ of the messenger singlet $X$
occurs when $k_1 < \lambda$  \cite{dnns} at the values,
\begin{eqnarray}
	\label{vx}
	| X |^2 ~&=&~ {1 \over k_1^2} ~{\lambda - k_1 \over 2 \lambda - k_1}
	~m_\phi^2 \\
	| \phi_+ |^2 = |\phi_- |^2 ~&=& ~ { 1 \over k_1^3} ~
	{ \lambda^2\over 2 \lambda - k_1} ~m_\phi^2,
\end{eqnarray}
where the $\phi_\pm$ fields are constrained to be equal by the messenger
hypercharge D-term.
The messenger quark and lepton superfields do not obtain {\it vev}s,
provided
$\lambda \ll k_{2,3}$ \cite{dnns}.
The $F_X$ expectation value is of order
\begin{equation}
	\label{fx}
	F_X  \simeq \lambda~\langle X\rangle^2 .
\end{equation}

This SUSY-breaking vacuum expectation value
of the gauge singlet
superfield $X$ splits the masses of the
superpartners in the messenger quark and lepton
supermultiplets. This splitting is further
transmitted to the standard model sector by
loops involving the
messenger quarks and leptons.
In this class of models, gaugino masses appear at one loop order,
while squark and slepton masses appear
at two-loop order. These soft SUSY-breaking parameters are approximately
\begin{equation}
	\label{msoft}
	\tilde{m} \simeq  {g^2\over 16 \pi^2} ~\lambda
	 \langle X\rangle ~\equiv~ {g^2\over 16 \pi^2} ~\Lambda,
\end{equation}
where $g$ is a standard model gauge coupling, and we have introduced the
dimensionful parameter $\Lambda$ as in Ref.~\cite{dn}. The scale of all
soft supersymmetry breaking parameters in the
low-energy theory is determined by $\Lambda$.
To obtain soft masses of order the electroweak scale,
$\tilde{m} \simeq 10^{2}$ GeV, we require $\Lambda \simeq 10^{4}$ GeV.

Finally, there is a coupling between the gauge singlet
messenger field $X$ and the Higgs doublets
\begin{equation}
	\label{muterm}
	W_{\mu} = \lambda' ~X H_U H_D ,
\end{equation}
 which generates a $B$ term in the scalar potential,
$B = m_{12}^2 H_U H_D$, and a $\mu$ term in the superpotential,
with parameters
\begin{eqnarray}
	\label{bandmu}
	m_{12} &=& \sqrt{\lambda' \lambda} ~\langle X\rangle
        = \sqrt{\lambda' \over \lambda} ~\Lambda  \nonumber \\
	\mu &=& \lambda' \langle X\rangle = {\lambda' \over \lambda} \Lambda .
\end{eqnarray}
A $\mu$ term of order $100$ GeV requires $\lambda'/\lambda \simeq 10^{-2}$,
which then determines the $B$ term to be of order $1$ TeV.
Notice that if one requires a $100$ GeV $B$ term, then the $\mu$ term becomes
unacceptably small for the simple prescription of Eq.~(\ref{muterm}).
This problem can be
overcome, as was done in Ref.~\cite{dnns}, by extending the Higgs sector with
the addition of extra gauge singlet
fields which also have nonrenormalizable interactions with the supercolor
sector.
Since a $B$ term parameter $m_{12}$ of order $1$ TeV is not ruled out
experimentally we will consider the simpler scenario of Eq.~(\ref{muterm}),
which yields the relations of Eq.~(\ref{bandmu}).

As we saw above, in order to avoid colour and electroweak-breaking
{\it vev}s of the
messenger quarks and leptons, we require $\lambda \ll k_{2,3} \simeq 1$.
On the other hand, obtaining  soft mass parameters $\tilde{m}$, $\mu$, and
$B$ of electroweak order of magnitude requires fine tuning of
the ratio $\lambda'/\lambda$.
In the following sections, we propose an explanation of this
hierarchy of couplings in the superpotentials,
Eqs.~(\ref{wsb}) and (\ref{muterm}),
by assuming that they arise from
higher dimension terms, constrained by a horizontal
anomalous $U(1)$ symmetry.
This symmetry, broken at some high energy scale
(e.g. the string scale, $M_S$), may also be used
to explain the fermion mass hierarchy. Thus the hierarchies of
the fermion masses,
as well as the correct order of magnitude of the soft supersymmetry
breaking terms
are explained by the same mechanism\footnote{Recently,
anomalous horizontal symmetries have been invoked to explain the
value of the $\mu$ term in hidden sector supergravity models
\cite{jain,nir}.}.

\section{String-Inspired Hierarchy }

In this section we will introduce the anomalous horizontal $U(1)_X$ symmetry,
review its application for
generating the fermion mass hierarchy, and extend it to explain the  order
of magnitude of the $\mu$ and $B$ terms in the low-energy
supersymmetry breaking
scenario. We assume the existence of a standard model gauge singlet
field, $\theta$,
which is charged under the $U(1)_X$ symmetry\footnote{In the next section we
will also consider a model with two standard model gauge singlet fields.}.
This field couples to the standard model
and messenger sector fields only through nonrenormalizable interactions,
suppressed by the string scale $M_S$.

\def\toM{\left({\theta\over M_S} \right)}

The total superpotential in which we will be interested
is the sum of the messenger sector, Eq.~(\ref{wsb}),
and the SUSY standard model superpotentials
\begin{equation}
	W = W_M + W_{SM},
\end{equation}
where
\begin{eqnarray}
	\label{WSB1}
	W_M ~= & & \xi_\phi \toM^{q_X + q_{\phi_+}
		+ q_{\phi_-}} X \phi_+ \phi_-
		+ \toM^{q_X} \left[ \xi_l X \bar{l}l
		+ \xi_q X \bar{q}q \right]
	        \nonumber \\
	&+& {\xi_X  \over 3} \toM^{3q_X} X^3
		+ \xi_H \toM^{q_H + q_X} X H_U H_D~.
\end{eqnarray}
The various powers of $\theta$  are determined by the $U(1)_X$
charge assignments, where  the $\theta$  field has charge $-1$.
The coefficients $\xi_\alpha$ are assumed to be numbers of order one
and $q_H = q_{H_U} + q_{H_D}$. The charge of $X$ is
$q_X = 1$ and  $\phi_\pm$ have charges $q_{\phi_+} = q_{\phi_-} = 1$.
With these charge assignments
we need $\xi_\phi < \xi_X$ in order to obtain the
required minimum, Eq.~(\ref{fx}).
We assume that the messenger quarks and leptons are in vectorlike
representations with respect to the
$U(1)_X$ symmetry as well, and therefore they do not contribute to
the various mixed anomalies considered below.
The standard model Yukawa couplings are generated by
\begin{equation}
	\label{WSM}
	W_{SM} = \xi^U_{ij} \toM^{p_{ij}^U} Q_i H_U u_j
		+ \xi^D_{ij} \toM^{p_{ij}^D} Q_i H_D d_j
		+ \xi^L_{ij} \toM^{p_{ij}^L} L_i H_D e_j,
\end{equation}
where $i,j = 1,2,3$ are generation indices.

As shown in Ref.~\cite{binetruy}, in order to explain the fermion
mass hierarchy,
the $U(1)_X$ symmetry must be anomalous. The anomaly is assumed
to be cancelled by the Green-Schwarz mechanism.
The  mixed anomaly coefficients for $U(1)_X$
and the standard model gauge groups are given by
\begin{eqnarray}
\label{anomalies}
	C_1 &=& {1\over 6}\left( 3\left[q_{H_U} + q_{H_D}\right]
		 + \sum_{i=1}^3 \left[q_{Q_i} + 8 q_{u_i}
			+ 2 q_{d_i} + 3 q_{L_i} + 6 q_{e_i} \right] \right),\\
	C_2 &=& {1\over 2} \left( q_{H_U} + q_{H_D}
		+ \sum_{i=1}^3 \left[ 3 q_{Q_i} + q_{L_i}\right]\right),\\
	C_3 &=& {1\over 2}\sum_{i=1}^3 \left(2 q_{Q_i} + q_{u_i} +
		q_{d_i} \right),\\
	C_{XXY} &=&  q_{H_U}^2 - q_{H_D}^2
			+ \sum_{i=1}^3 \left[ q_{Q_i}^2 - 2 q_{u_i}^2
						+ q_{d_i}^2 - q_{L_i}^2
						+ q_{e_i}^2 \right].
\end{eqnarray}
Here $C_1, C_2, C_3$ are the coefficients of the mixed
$U(1)_X U(1)_Y^2$, $U(1)_X SU(2)^2$, $U(1)_X SU(3)^2$ anomalies,
respectively, and
$C_{XXY}$ is the $U(1)_X^2 U(1)_Y$ anomaly. The Green-Schwarz
mechanism \cite{gs} for anomaly cancellation requires that
\begin{equation}
{C_1\over k_1}~ =~{ C_2\over k_2}~ = ~{C_3\over k_3}~ =
 ~{C_X\over k_X}~ =~ {C_{grav}\over
 k_{grav}} ~,
\label{condition1}
\end{equation}
where $k_i$ denotes the Kac-Moody levels of the corresponding
gauge algebras, $C_{grav} = (1/24) {\rm Tr} Q$ (the sum
of all the $U(1)_X$ charges in the model), and $C_X$ is the
coefficient of the $U(1)_X^3$ anomaly.
The $U(1)_X^2 U(1)_Y$ anomaly cannot be cancelled by the Green-Schwarz
mechanism. We will therefore only
consider charge assignments for which
\begin{equation}
	C_{XXY} ~=~ 0~.
	\label{condition2}
\end{equation}
We will consider Kac-Moody levels with the values $k_1 = 5/3, k_2 = k_3
= k_{grav} = 1$ and $C_X$ will not be used because there may be other
$U(1)_X$-charged fields in the theory
which are singlets under the standard model gauge groups.
Similarly we do not consider the mixed anomalies with the supercolor sector
gauge group(s), but simply assume that these can be made consistent with
the Green-Schwarz mechanism by means of appropriate charge assignments.
The gravitational anomaly $C_{grav}$ also depends upon the
$U(1)_X$ charges of any additional fields, however by the Green-Schwarz
mechanism it is proportional to $C_3$. This will be important
below for determining the {\it vev} of $\theta$ .

String loop effects induce a Fayet-Iliopoulos term for the anomalous
$U(1)_X$ symmetry~\cite{dws}. The $U(1)_X$ D-term is then given by
\begin{equation}
\label{dterm}
D = {g_S^2 M_S^2 \over 192 \pi^2}~ {\rm Tr} Q ~+ ~\sum_i q_i ~|\phi_i|^2~,
\end{equation}
where $g_S$ and $M_S$ are the string coupling and string scale,
respectively, and
the sum is over all fields carrying $U(1)_X$ charge. Requiring
unbroken supersymmetry at the string scale, with our choice of
charge for $\theta$, we find that the
{\it vev} of $\theta$ is then given by
\begin{equation}
	{\langle\theta\rangle \over M_S} =
	\sqrt{ {g_S^2\over 192 \pi^2} ~{\rm Tr} Q }
  =   \sqrt{ {g_S^2\over 8 \pi^2}~ C_3 }~,
\end{equation}
where the last equality holds for our choice of Kac-Moody levels.
Using the tree-level relation $1/g_S^2 = k_{GUT}/g^2_{GUT}$
at the unification scale%
\footnote{We assume string-type unification. For a discussion of the
phenomenology of this model in an $SU(5)$-unified framework,
see Ref. \cite{hitoshi}.}, and taking $k_{GUT} = 1$ and the typical value
$g_{GUT}^2/4 \pi \simeq 1/25$, we obtain
\begin{equation}
	\label{theta}
	{\langle\theta\rangle \over M_S} \simeq 0.08 \sqrt{C_3} ~.
\end{equation}
In the next section we will study
two specific examples in which the hierarchies are parameterised
by various powers of this ratio, with the powers
determined from $U(1)_X$ charge
assignments of the various fields.

\section{Examples of $U(1)_X$ models}

\subsection{One $\theta$-field case}

Consider the case of one $\theta$ field which couples to all terms
in the superpotential. In order to determine the allowed couplings
one can perform a computer search,
along the lines of Ref.~\cite{rz}, for charges that obey the
conditions of Eqs.~(\ref{condition1}) and (\ref{condition2})
and that allow for $q_H > 0$,
which leads to an acceptable fermion mass hierarchy.
However, in this paper we
will restrict ourselves to giving an existence proof for
simple analytic solutions.
For left-right symmetric $U(1)_X$ charge assignments, one can
analytically solve the
$U(1)_X^2 U(1)_Y$-anomaly condition (\ref{condition2}) \cite{jain}.
However, in the one $\theta$ field case, it was
shown in Ref.~\cite{jain} that the fermion mass hierarchy cannot be
explained by assuming left-right symmetric $U(1)_X$ charge assignments.
A solution which exhibits the required asymmetry is given by the following
quark and lepton $U(1)_X$ charge assignments
\begin{equation}
	\begin{array}{cccccc}
	\;\;i\;\; & \;\;\;q_{Q_i}\;\; \;&\;\;
		\;q_{u_i}\;\;\; & \;\;\; q_{d_i}\;
	 \; \;&\; \;\; q_{L_i}\;\; \;&\; \;\; q_{e_i}\;\;\; \\
	&&&&&\\
	1 & {33\over 40} & {403\over 240} & {401\over 240} & {143\over 120}
	  & {73\over 240}  \\
	&&&&&\\
	2 & {13\over 40} & {43\over 240} & {161\over 240} & {83\over 120}
	  & -{47\over 240} \\
	&&&&&\\
	3& -{27\over 40} & -{197\over 240} & {401\over 240} & {83\over 120}
	 & -{47\over 240} \\
	\end{array}
	\label{onetheta}
\end{equation}
where the two Higgs superfields carry charges $q_{H_u}=359/240$ and
$q_{H_d}=121/240$, while $X$ and $\theta$ have charges $q_X=1/2$ and
$q_\theta=-1/2$. In addition one needs to charge the
$\phi_\pm$ fields,
$q_{\phi_\pm}=1/2$ so that the correct SUSY-breaking minimum is achieved
(\ref{vx}). The charge assignments (\ref{onetheta}) lead to $C_3=3$ and
from (\ref{theta}) we obtain
$\epsilon \equiv \langle\theta\rangle/M_S \simeq 0.1$.

As noted earlier, the relation (\ref{theta}) depends on the choice
of Kac-Moody levels, which we have fixed.
In addition, there are couplings in the superpotential, $\xi_\alpha$,
of order one. Here we assume that the values of
these couplings (and $k_{grav}$)
 are such that $\epsilon \simeq 0.2$ in order to agree
with previous parameterizations of the fermion mass hierarchies
\cite{binetruy,dudas,jain}.
The important point to note is that the correct order of magnitude for
$\epsilon$ is obtained.

Thus we obtain $\lambda \simeq \epsilon^3, \lambda' \simeq \epsilon^5$,
and a ratio
$\lambda'/\lambda \simeq\epsilon^2 \simeq10^{-2}$, as required earlier
in Sec.~2.
The Yukawa coupling matrices for the quarks are determined to be
\begin{equation}
	Y^u\simeq\left ( \begin{array}{ccc}
	      \epsilon^8 & \epsilon^5 & \epsilon^3 \\
	      \epsilon^7 & \epsilon^4 & \epsilon^2 \\
	      \epsilon^5 & \epsilon^2 & 1 \end{array} \right ) , \quad
	Y^d\simeq \left ( \begin{array}{ccc}
	      \epsilon^6 & \epsilon^4 & \epsilon^6 \\
	      \epsilon^5 & \epsilon^3 & \epsilon^5 \\
	      \epsilon^3 & \epsilon^1 & \epsilon^3 \end{array} \right ),
	\label{qyukawaone}
\end{equation}
while the lepton Yukawa coupling matrix is given by
\begin{equation}
	Y^l\simeq\left ( \begin{array}{ccc}
	      \epsilon^4 & \epsilon^3 & \epsilon^3 \\
	      \epsilon^3 & \epsilon^2 & \epsilon^2 \\
	      \epsilon^3 & \epsilon^2 & \epsilon^2 \end{array} \right ) .
	\label{lyukawaone}
\end{equation}
These Yukawa coupling matrices give the correct phenomenological mass
ratios \cite{RRR} for the quarks
\begin{equation}
	m_u/m_c \simeq \epsilon^4, \quad m_c/m_t \simeq \epsilon^4, \quad
	m_d/m_s \simeq \epsilon^2, \quad m_s/m_b \simeq \epsilon^2
	\label{quarkratios}
\end{equation}
and the leptons
\begin{equation}
	m_e/m_\mu \simeq \epsilon^2, \quad m_\mu/m_\tau \simeq \epsilon^2 ~.
	\label{leptonratios}
\end{equation}
It should be pointed out that a moderate amount of tuning
$(\cal O(\epsilon))$ of the $\xi_{ij}$ coefficients is implicit in
order to obtain these mass ratios,
since otherwise the rank of these matrices would be generically
nonmaximal. This is typically the case for these types of models.

Note that any phase of the $\theta$ field can be rotated
into the $\Theta$-angles of the corresponding gauge
groups%
\footnote{For supercolor sectors like the ones considered
	in Ref.~\cite{dnns} (with a single nonperturbative term in
	the superpotential) this phase will dynamically relax to zero.
}
by field redefinitions of all $U(1)_X$-charged fields (including those in the
supercolor sector). Thus in order to
incorporate CP-violation, it is necessary to assume that the coefficients
$\xi_{ij}$ in Eq.~(\ref{WSM}) are complex.
In the next example we will consider
a solution with two $\theta$-fields which can have symmetric charge
assigments and nontrivial phases in the Yukawa couplings.

\subsection{Two $\theta$-field case}

Consider a solution with two $\theta$-fields,
$\theta$ and $\theta'$, with $U(1)_X$ charges $q_\theta=-1/2$ and
$q_{\theta^\prime}=-3/4$.
We will use the analytic parametrization presented in Ref.~\cite{jain}
for the  solutions of the Green-Schwarz anomaly cancellation conditions
(\ref{condition1}), (\ref{condition2}) with left-right symmetric charge
assignments.
The most general superpotential allowed by the symmetries is now given
by (\ref{WSB1}) and (\ref{WSM}), where either $\theta$ or $\theta'$
(or a product of both) can appear in place of the single field $\theta$,
in a way consistent with the $U(1)_X$ charge  assignments.
We will impose a $Z_2\times Z_2$ discrete symmetry permitting only an even
number of $\theta$ and $\theta'$ fields.

The solution which we describe below has a coefficient of the mixed
$U(1)_X SU(3)^2$ anomaly $C_3 = 4$.
Generalizing Eq.~(\ref{theta}) to the case of two fields,
we find that the D-term vanishes for
\begin{equation}
	\label{twotheta}
	{1\over 2} {\langle\theta\rangle \over M_S}^2 +
	{3\over 4} {\langle\theta'\rangle \over M_S}^2 ~\simeq ~(0.2)^2 .
\end{equation}
We will assume that the {\it vev}s of the two fields are of the
same order of magnitude,
$\epsilon =\langle\theta\rangle/M_S \simeq
\langle\theta'\rangle/M_S \simeq 0.2 $.

The charges of the quark and lepton superfields under the $U(1)_X$
symmetry are as follows
\begin{equation}
	\begin{array}{cccccc}
	\;\;i\;\; & \;\;\;q_{Q_i}\;\; \;&\;\;
		\;q_{u_i}\;\;\; & \;\;\; q_{d_i}\;
	 \; \;&\; \;\; q_{L_i}\;\; \;&\; \;\; q_{e_i}\;\;\; \\
	&&&&&\\
	1 & {29\over 15} & {29\over 15} & {11\over 5} &
		{9\over 20} & {71\over 60}  \\
	&&&&&\\
	2 & {13\over 30} & {13\over 30} & {7\over 10} &
		 -{3\over 10} & {13\over 30} \\
	&&&&&\\
	3& -{17\over 30} & -{17\over 30} & -{3\over 10} & -{11\over 20} &
	{11\over 60} .\\
	\end{array}
	\label{charges}
\end{equation}
The two Higgs fields have charges $q_{H_U} = 17/15$ and $q_{H_D} = 28/15$.
The sum of
their charges is therefore $q_H = 3$ and allows for the generation of
a $\mu$ term from the
superpotential. The messenger field $X$ has charge $q_X=1$. With these
charge assignments, we obtain
$\lambda \simeq \epsilon^4, \lambda' \simeq \epsilon^6$, and a ratio
$\lambda'/\lambda \simeq\epsilon^2 \simeq10^{-2}$, as discussed in Sec.~2.

The charge assignments (\ref{charges}) lead to the following quark
Yukawa matrices,
\begin{equation}
	Y^u\simeq\left ( \begin{array}{ccc}
	      \epsilon^8 & \epsilon^6 & \epsilon^4 \\
	      \epsilon^6 & \epsilon^4 & \epsilon^2 \\
	      \epsilon^4 & \epsilon^2 & 1 \end{array} \right ) , \quad
	Y^d\simeq  \epsilon^2 \left ( \begin{array}{ccc}
	      \epsilon^6 & \epsilon^4 & \epsilon^4 \\
	      \epsilon^4 & \epsilon^2 & \epsilon^2 \\
	      \epsilon^4 & \epsilon^2 & 1 \end{array} \right ).
	\label{qyukawa}
\end{equation}
Similarly the lepton Yukawa matrix is given by
\begin{equation}
	Y^l\simeq\left ( \begin{array}{ccc}
	      \epsilon^6 & 0 & \epsilon^4 \\
	          0 & \epsilon^4 & 0 \\
	      \epsilon^4 & 0 & \epsilon^2 \end{array} \right ) .
	\label{lyukawa}
\end{equation}
The above Yukawa matrices give the correct mass ratios for the quarks
(\ref{quarkratios}) and leptons (\ref{leptonratios}).
Note that, in order to reproduce the correct mass ratios for the down
type quarks, $Y^d_{22}$,$Y^d_{23}$ and $Y^d_{32}$ must be
tuned slightly. The required tuning of the $\xi$'s is at most of order
$\epsilon$, so this is acceptable. Also, we presume there is some
freedom in the adjustment of $\langle\theta\rangle / \langle\theta'\rangle$,
which affects the $Y^\alpha_{ij}$ as well at order one.

Finally we would like to note the possibility of spontaneous CP
violation in models with anomalous horizontal symmetries with
more than one $\theta$ field. In the example above with two
fields, $\theta$ and $\theta'$, one linear combination
of the two phases%
\footnote{To be precise, we should note
	that one linear combination of the string
	axion and the phases of $\theta$ and $\theta'$ becomes the longitudinal
	component of the $U(1)_X$ gauge field, which obtains a
	mass of order the string scale \cite{dws}.
	Since possible kinetic-term mixing is not important
	for our considerations, we can parameterize
	the two remaining linear combinations of phases in the
	low energy theory
	as the phases of $\theta$ and $\theta'$.}
can be rotated
away into the $\Theta$ angles of the corresponding gauge groups,
and this combination is expected to
relax to zero due to nonperturbative effects in the supercolor sector.
The other linear combination of the two phases can be rotated away from the
SUSY-breaking parameters in the
standard model (by field redefinitions of the messenger quarks and leptons
and the two Higgs doublets) and appears in the Yukawa couplings and the QCD
$\Theta$ angle. However, this phase is also
expected to appear in the supercolor-sector interactions; in order to
dynamically determine its {\it vev},
one needs to specify in more detail the supersymmetry breaking
dynamics. This possible source of CP violation requires further investigation.

\section{Conclusion}

Models of low-energy supersymmetry breaking offer the hope
for new physics at intermediate energy scales. It is therefore worthwhile
to understand what is required in order to make these models
realistic in detail. We have indicated here one option for
building realistic models of this type; other options have
also been studied in the literature\cite{dns,dnns}.
We have shown that the fermion mass hierarchy and the values of the
$\mu$ and $B$ terms in supersymmetric extensions of the standard model
with low-energy supersymmetry breaking can have a common origin.
The values of all Yukawa couplings are determined by a spontaneously broken,
anomalous $U(1)$ horizontal symmetry, the anomaly being cancelled by the
Green-Schwarz mechanism. As an existence proof we have constructed
explicit models of this type with both one and two $\theta$-fields.

\bigskip

\section{Acknowledgements}

We are grateful to the Aspen Center for Physics, where this
work was begun. We would like to thank J. Harvey, M. Robinson,
and J. Ziabicki for useful conversations. TG also acknowledges
discussions with M. Einhorn and S. Martin, and GJ thanks J. Schechter
for discussions about mass matrices.
TG was supported by the DOE at the University of Michigan.
GJ was supported by the DOE under contract DE-FG02-85-ER40231
at Syracuse University.
EP was supported by a Robert R. McCormick Fellowship and by
the DOE under contract  DF-FC02-94-ER40818.

\vfil\eject


\begin{references}

\bibitem{dnns}M. Dine, A.E. Nelson, Y. Nir, and Y. Shirman, hep-ph/9507378;
A.E. Nelson, hep-ph/9511218.

\bibitem{dn}M. Dine and A.E. Nelson, Phys. Rev. {\bf D48} (1993) 1277.

\bibitem{dns}M. Dine, A.E. Nelson and Y. Shirman, Phys. Rev. {\bf D51}
(1995) 1362.

\bibitem{ibanez}L. Ibanez and G.G. Ross, Phys. Lett. {\bf B332} (1994) 100.

\bibitem{binetruy}P. Binetruy and P. Ramond, Phys. Lett. {\bf B350} (1995) 49.

\bibitem{dudas}E. Dudas, S. Pokorski and C.A. Savoy, Phys. Lett. {\bf B356}
(1995) 45.

\bibitem{jain}V. Jain and R. Schrock, Phys. Lett. {\bf B352} (1995) 83;
V. Jain and R. Schrock, hep-ph/9507238.

\bibitem{nir}Y. Nir, Phys. Lett. {\bf B354} (1995) 107.

\bibitem{rz}M. Robinson and J. Ziabicki, EFI-95-50, hep-ph/9508322.

\bibitem{fn}C. Froggatt and H. Nielsen, Nucl. Phys. {\bf B147} (1979) 277.

\bibitem{gs}M. Green and J. Schwarz, Phys. Lett. {\bf B149} (1984) 117.

\bibitem{dws}M. Dine, N. Seiberg and E. Witten, Nucl. Phys. {\bf B289}
(1987) 589;
J.J. Attick, L. Dixon and A. Sen, Nucl. Phys. {\bf B292} (1987) 109;
M. Dine, I. Ichinose and N. Seiberg, Nucl. Phys. {\bf B293} (1987) 253.

\bibitem{hitoshi}C.D. Carone and H. Murayama, LBL-37810, hep-ph/9510219.

\bibitem{RRR}P. Ramond, R.G. Roberts and G.G. Ross, Nucl. Phys.
{\bf B406} (1993) 19.

\end{references}
\end{document}